\documentstyle[sprocl,epsfig]{article}

\def\Journal#1#2#3#4{{#1} {\bf #2}, #3 (#4)}

\def\MN{{\em MNRAS}}
\def\ApJ{{\em ApJ}}
\def\AA{{\em A\&A}}

\def\be{\begin{equation}}
\def\ee{\end{equation}}
\def\bea{\begin{eqnarray}}
\def\eea{\end{eqnarray}}

\newbox\grsign \setbox\grsign=\hbox{$>$} \newdimen\grdimen \grdimen=\ht\grsign
\newbox\simlessbox \newbox\simgreatbox \newbox\simpropbox
\setbox\simgreatbox=\hbox{\raise.5ex\hbox{$>$}\llap
     {\lower.5ex\hbox{$\sim$}}}\ht1=\grdimen\dp1=0pt
\setbox\simlessbox=\hbox{\raise.5ex\hbox{$<$}\llap
     {\lower.5ex\hbox{$\sim$}}}\ht2=\grdimen\dp2=0pt
\setbox\simpropbox=\hbox{\raise.5ex\hbox{$\propto$}\llap
     {\lower.5ex\hbox{$\sim$}}}\ht2=\grdimen\dp2=0pt
\def\simgt{\mathrel{\copy\simgreatbox}}

\def\jcr{j_{\rm cr}}
\def\lcr{j_{\rm cr}}
\def\jK{j_{\rm K}}
\def\OmK{\Omega_{\rm K}}
\def\Tbb{T_{\rm bb}}
\def\tT{\tau_{\rm T}}
\def\sT{\sigma_{\rm T}}
\def\TC{T_{\rm C}}
\def\Tesc{T_{\rm esc}}
\def\RC{R_{\rm C}}
\def\rcr{r_{\rm cr}}
\def\tC{t_{\rm C}}
\def\LE{L_{\rm E}}
\def\dd{{\rm d}}
\def\dM{\dot{M}}

\def\thi{\theta_\infty}

\def\tep{t_{\rm Coul}}
\def\vff{v_{\rm ff}}

\def\bl{\bar{l}}
\def\Ed{E_{\rm diss}}

\begin{document}

\title{MODES OF ACCRETION IN X-RAY SOURCES}

\author{A. M. BELOBORODOV}

\address{Canadian Institute for Theoretical Astrophysics\\
University of Toronto, 60 St. George Street \\
Toronto, M5S 3H8 Ontario, CANADA}

\maketitle\abstracts{ Three classical modes of accretion are briefly discussed: 
wind-fed, spherical, and disk. The three modes are illustrated with the
mass transfer onto black holes in high-mass X-ray binaries. Then a new regime 
of mini-disk accretion is described and it is argued that observed wind-fed 
X-ray sources are likely to accrete in this regime. Switching from 
one accretion mode to another can cause the observed spectral state transitions.
}

\section{Introduction}
Accretion became an important subject of astrophysics 3 decades ago when it was 
realized to feed the brightest X-ray sources in the sky --- X-ray binaries. 
These sources are associated with the most compact astrophysical
objects: neutron stars and black holes. Accretion onto such objects has 
four major features: (1) It releases
energy in a relativistically deep potential well, $\phi\sim -c^2$, and can have
a high radiative efficiency. (2) It produces radiation in a small region 
of radius $r$ comparable to the Schwarzschild radius 
$r_g=2GM/c^2=0.3\times 10^6(M/M_\odot)$ where $M$ is the object mass. 
(3) The observed luminosities $L$ are often comparable to the Eddington limit 
$\LE=2\pi r_g m_p c^3/\sT\approx 1.3\times 10^{38}(M/M_\odot)$~erg/s. 
(4) The high $L$ combined with the small size imply a high temperature of 
the source: a lower bound on the temperature can be estimated assuming blackbody 
emission, $kT_{\rm bb}\approx (L/\sigma r^2)^{1/4}\sim$~keV. This 
emission is in the X-ray band. 

There are three textbook regimes of accretion:
wind-fed (Hoyle \& Lyttleton~\cite{HL}; Bondi \& Hoyle~\cite{BH}),
spherical (Bondi~\cite{Bondi}), and disk (Shakura \& Sunyaev~\cite{SS}).
In all cases gas falls onto the accretor and releases its 
gravitational energy, yet the specific accretion mechanisms and the flow 
patterns differ. Here we briefly discuss these classical modes and 
illustrate them with black-hole accretion in high-mass X-ray binaries.
Then we describe a new regime of mini-disk accretion.


\section{Wind-fed X-ray Binaries}
Mass transfer in an X-ray binary strongly depends on the  
type of the normal companion (donor). If the donor is massive (OB star) it emits 
a powerful wind which is partially captured by the compact companion.
Thus, accretion occurs even if the donor does not fill its Roche lobe.
We will focus on the wind-fed accretion here because it provides nice illustrations
of different accretion modes. Also, historically, the first accretion scenario
(Bondi-Hoyle-Lyttleton) was wind-fed.

The wind material is captured from a cylinder around the line that connects
the two companions. It is called the accretion cylinder and its radius is
\be
R_a=\frac{2GM}{w^2}\approx 3\times 10^{10}
    \left(\frac{M}{M_\odot}\right)\left(\frac{w}{10^8}\right)^{-2}{\rm ~cm},
\ee
where $w$ is the wind velocity and $M$ is the mass of the accretor.
Winds from massive OB stars have $w\approx 10^8$~cm~s$^{-1}$, 
approximately equal to the escape velocity from the star surface.
The separation of the observed X-ray binaries, $A\sim 10^{12}$~cm, is 
much larger than $R_a$. This has two consequences: (1) only a small fraction 
$\sim (R_a/A)^2$ of the wind is captured and (2) the flow inside the accretion 
cylinder is nearly plane-parallel before it gets trapped by the gravitational 
field of the accretor. 

The wind crosses the distance $A$ and accretes on a timescale $A/w\sim 10^4$~s.
It is much shorter than the orbital period of the binary, $P\sim 10^5-10^6$~s
and one could expect that the orbital rotation weakly affects the accretion flow;
in particular, the net angular momentum $\bar{l}$ of the trapped gas must be close 
to zero. In fact, even a slow orbital rotation and a corresponding small 
$\bar{l}\neq 0$ can strongly impact the mechanism of accretion:
the trapped gas falls many decades in radius onto the compact object 
($R_a/r_g\approx 10^5$) and the approximation $\bar{l}=0$, which is good at 
$R\sim R_a$, can fail before the gas reaches $r_g$. A very small $\bar{l}\sim r_g c$ 
would be sufficient for a disk formation near the compact object.

The trapped $\bl$ was estimated and compared with $r_gc$ by Illarionov \& 
Sunyaev~\cite{IS}  and Shapiro \& Lightman~\cite{ShapL}. Consider the accretion 
cylinder as a sequence of moving pancakes of radius $R_a$ (perpendicular slices 
of the cylinder). When viewed from the accretor, the pancakes have the orbital 
angular velocity $\Omega=2\pi/P$ and their specific angular momentum is 
$\bl=(1/4)\Omega R_a^2$. Using equation~(1) one gets
$\bl/(r_g c)\approx (P/{\rm day})^{-1}(M/{\rm M}_\odot)(w/10^8)^{-4}$.
Three X-ray binaries are classified as massive ones with black hole
companions: Cyg~X-1, LMC~X-1, and LMC~X-3 (see Tanaka \& Lewin~\cite{TL}).
They have orbital periods $P=5.6$, 4.2, and 1.7~d, respectively, and masses
$M$ of order of $10 M_\odot$.
One thus obtains $\bl\sim r_gc$ for these objects, which is marginal for 
formation of a small-scale disk. Whether the disk forms or not depends on the 
precise value of $\bl$ which is difficult to calculate because the 
details of wind accretion are not fully understood (note the strong dependence
of $\bar{l}$ on the wind velocity, $w$). 

At distances $R\gg r_g$ from the accretor one can assume $\bl=0$ and we first
discuss models with zero net angular momentum (Sect.~3 and 4). 

\begin{figure}[t]
\center{\epsfig{file=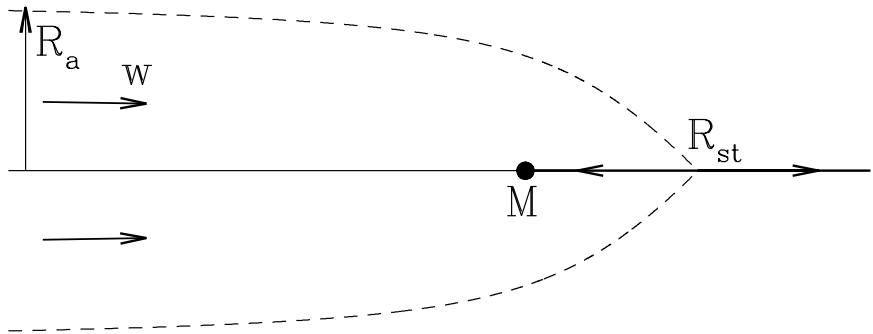,height=5cm}}
\caption{ Bondi-Hoyle-Lyttleton accretion. An initially
plane-parallel flow is deflected by the accretor and focused onto the axis
of symmetry (the accretion line). Further accretion proceeds along the line 
inside a stagnation radius $R_{\rm st}$.
}
\end{figure}


\section{Bondi-Hoyle-Lyttleton Accretion}
The original model of Bondi \& Hoyle~\cite{BH} envisions an initially plane-parallel 
flow interacting with an accretor of mass $M$ (Fig.~1). The accretor deflects the gas 
streamlines from the straight lines and captures a part of the flow. The model
neglects the gas pressure, so the streamlines follow ballistic (free-fall) 
trajectories of test particles in the gravitational field of the accretor. 
The problem has a symmetry axis and the deflected trajectories intersect 
on the axis behind the accretor. This one-dimensional caustic (called 
the accretion line) is assumed to be sticky.
The gas in the caustic has a velocity $u^r$ which differ from the 
the radial component $\hat{u}^r$ of the free fall that impinges on the 
caustic. The gas motion in the caustic obeys the momentum conservation law, 
\be
\label{eq:bhl}
 \frac{\dd u^r}{\dd r}=\frac{\dd\dM}{\dd r}\frac{(\hat{u}^r-u^r)}{\dM(r)}
-\frac{GM}{r^2u^r}.
\ee
Here $\dM(r)$ is the flux of mass in the caustic, and $\dd \dM/\dd r$ is its
feeding rate, which is easily calculated given the ballistic trajectories of
the infall. The infall momentum is directed away from the 
accretor and its inertia competes with gravity. Gravity dominates at small 
$r$ and here $u^r<0$ (gas flows in) while at large radii $u^r>0$ (gas flows out). 
Hence there must be a stagnation point on the line at a radius 
$R_{\rm st}\sim R_a$. The $R_{\rm st}$ is not well defined in the model --- 
infinite number of steady solutions with different $R_{\rm st}$ exist.
For a given solution $u^r(r)$ one easily calculates the dissipated energy:
$\dd\Ed/\dd r=(\dd\dM/\dd r)[\hat{u}_\perp^2+(\hat{u}^r-u^r)^2]/2$ where
$\hat{u}_\perp$ is the non-radial velocity component that is canceled where 
the ballistic flow reaches the caustic. 

This simple and beautiful model does not, however, apply to real X-ray sources.
The model is based on the assumption that the colliding gas on the 
accretion line immediately emits the dissipated energy. In fact, the 
radiative losses are small on the $R_a/w$ timescale and the stored heat creates 
a high pressure that forces the shocked gas to expand away from the caustic.
As a result a steady bow shock sets in at a distance $\sim R_a$ in front of the 
accretor. 
Hunt~\cite{Hunt} first studied the subsonic gas dynamics behind the shock and showed 
that a spherically symmetric inflow forms at distances $R<R_a$ from the accretor.
Subsequent numerical simulations (e.g. Blondin {\it et al.}~\cite{Blond}; 
Ruffert~\cite{Ruf97,Ruf99}) 
showed that the flow is unstable and variable if the bow shock is strong, 
i.e. if its Mach number ${\cal M}=w/c_s$ is well above unity 
[here $c_s=(10kT/3m_p)^{1/2}$ is the upstream sound speed]. In the case of a modest 
${\cal M}\simgt 1$ the variations are weak and the flow is approximately laminar.
This is the most likely case if accretion occurs in the radiation field of a
luminous X-ray source which preheats the upstream flow (Illarionov \& 
Beloborodov~\cite{IB}). 
The laminar transformation of an initially plane-parallel flow into a 
spherical infall is schematically shown in Fig.~2.

\begin{figure}[t]
\center{\epsfig{file=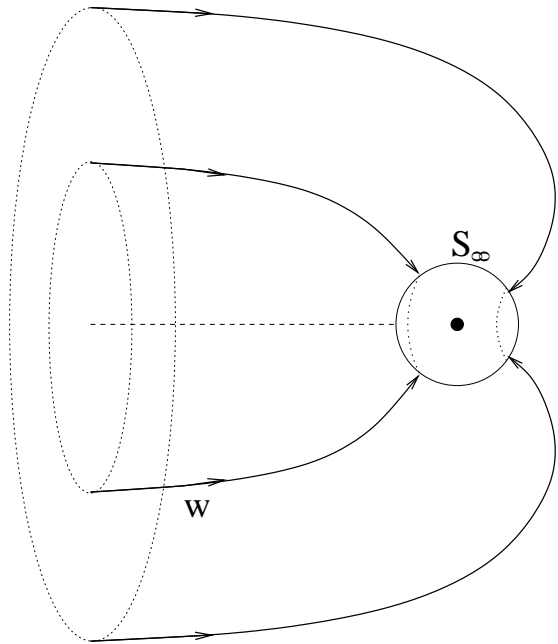,height=9cm}}
\caption{ Transformation of a plane-parallel flow into a spherical infall.
$S_\infty$ denotes the sphere of radius $R\sim\RC$ (eq.~6).
Inside $S_\infty$, the flow is Compton cooled and falling freely.
}
\end{figure}


\section{Spherical Accretion}
The first detailed study of 
spherically symmetric (radial) accretion was done by Bondi~\cite{Bondi}.
The steady gas dynamics is governed by simple equations of
baryon and momentum conservation,
\be
\label{eq:spher}
  4\pi R^2\rho u=\dM={\rm const}, \qquad
  u\frac{\dd u}{\dd R}=-\frac{1}{\rho}\frac{\dd p}{\dd R}-\frac{GM}{R^2}.
\ee
The problem has three unknowns: baryon density $\rho$, pressure $p$, and 
accretion velocity $u$. Hence one more equation is needed. 
In the Bondi model the set of equations is closed by assuming a polytropic 
equation of state $p=K\rho^\Gamma$. One can then find a family of solutions
specified by an accretion rate $\dM$ and an inner boundary condition. The inner 
boundary differs for 
neutron stars and black holes; in the latter case the flow must pass through 
a sonic point since the gas supersonically falls in at the event horizon.
The standard theory of polytropic spherical accretion is 
explained in details in textbooks (see e.g. Shapiro \& Teukolsky~\cite{ShapT}; 
Frank, King \& Raine~\cite{FKR}).

This theory, however, does not apply to luminous X-ray sources, 
as was pointed out by Zel'dovich \& Shakura~\cite{ZS}. The accreting gas exchanges 
energy with the radiation field of the source through Compton scattering and the 
effective $\Gamma\neq const$. Compton scattering dominates over other radiative 
processes and the thermal balance of the accreting gas reads 
\be
  \frac{\dd T}{\dd t}=-\frac{2}{3}T\,{\rm div}\,{\bf v}
                      +\frac{T_{\rm C}-T}{t_{\rm C}}.
\ee
The first term on the right-hand side is the compressive heating, and the second
term describes the energy exchange with the radiation on the Compton timescale
$\tC=3\pi m_ec^2 R^2/(\sT L)$. 
The $\TC$ is the Compton temperature of the radiation field.
If the radiation scatters off a plasma with $T>\TC$, the photons on average gain 
energy i.e. cool the plasma, while at $T<\TC$ the photons loose energy (because 
of the quantum recoil effect in Compton scattering) and heat the plasma. The 
$\TC$ is defined so that a plasma with $T=\TC$ is in energy equilibrium with the 
radiation field. It is determined by the shape of the radiation spectrum $F_\nu$, 
\be
   k\TC=\frac{\int F_\nu h\nu\dd\nu}{4\int F_\nu \dd\nu}.
\ee
In hard X-ray sources $\TC\sim 10^8$~K.

The temperature of the wind far from the accretor is relatively low,
$T_0\sim 10^5{\rm ~K}\ll\TC$, and where the flow approaches the X-ray source it is 
heated by Compton scattering. Inside $R_a$, just where the spherical inflow 
is forming, the gas is still heated, so that the effective $\Gamma>5/3$ and 
changes with $R$.
Compton heating strongly affects the flow, breaks the spherical symmetry,
and leads to formation of an inflow-outflow pattern of accretion 
(Ostriker {\it et al.}~\cite{Ostr}; Illarionov \& Kompaneets~\cite{IK}; 
Igumenshchev, Illarionov, \& Kompaneets~\cite{IIK}).
The situation changes inward of the so-called Compton radius 
\be
  \RC=\frac{GMm_p}{5k\TC}\approx 0.1 R_a.
\ee
Here the escape temperature $\Tesc(R)=GMm_p/(5kR)$ exceeds $\TC$ and a stable 
spherical inflow 
forms with ${\cal M}>1$. The plasma is now cooled by the X-rays rather than 
heated. The cooling timescale $\tC$ gets shorter than the compressive heating 
timescale $R/v$, and the gas is cooled well below $\Tesc$.
It implies that the pressure term in the momentum 
equation~(\ref{eq:spher}) is small compared to the gravitational term and gas
falls almost freely, with a high Mach number. Magnetic fields trapped from the
donor wind are amplified and dissipated in the converging infall (Bisnovatyi-Kogan 
\& Ruzmaikin~\cite{Bisn}; M\'esz\'aros~\cite{Mesz}). Their energy should not,
however, 
exceed the gas thermal energy, and therefore the fields should not affect the 
free-fall. 

The Compton-cooled radial free-fall at $R\ll \RC$ emits little radiation.
Its low pressure implies that the compressive heating is weak and the gravitational 
energy of the gas transforms largerly into the ballistic kinetic energy. 
If the compact object is a black hole, this energy is swallowed along with the 
flow, leading to a low radiative efficiency of accretion.
This regime can hardly take place in the observed bright sources like Cyg~X-1.
An appropriate mode of accretion for these sources should be able to dissipate 
efficiently the infall energy. Dissipation can naturally occur if the gas
has a non-zero angular momentum and forms a disk before plunging
into the event horizon. We now turn to accretion with non-zero angular momentum.


\section{Viscous Disk}
Accretion with angular momentum $l$ is stopped by the centrifugal
barrier at a radius $r=l^2/GM$. Here the gas forms a ring that rotates at 
about Keplerian angular velocity $\OmK=(GM/r^3)^{1/2}$. In this section we 
focus on models with $l\gg r_gc$. This is the condition of
the standard disk model where steady accretion is possible only if some mechanism 
redistributes angular momentum and allows the gas to spiral toward the centre. 
A plausible viscosity mechanism 
is provided by MHD turbulence that develops due to the differential character 
of Keplerian rotation (see Balbus \& Hawley~\cite{Balbus} for a review). 
Viscosity 
dissipates the orbital energy of the rotating gas and the gas diffuses to more 
bound circular orbits. A steady accretion disk dissipates energy with a rate 
$F_{\rm diss}$ per unit area,
\be
\label{eq:flux}
  F_{\rm diss}(r)=\frac{3\dM}{4\pi}\OmK^2 S,
\ee
where $S(r)$ is a numerical factor determined by the inner boundary condition.
For a Schwarzschild black hole $S=1-(3r_g/R)^{1/2}$ so that $S=0$ at the inner 
boundary of the viscous disk, $r=3r_g$, where the gas starts to fall freely into 
the hole. This standard disk model was successfully applied to X-ray binaries
and active galactic nuclei (AGN), the latter being just a scaled version of 
the standard model that corresponds to a massive black hole, $M=10^6-10^9M_\odot$.

If the dissipated energy is immediately radiated away, the emerging radiation 
flux from the disk surface is $F(r)=F_{\rm diss}(r)/2$ (the factor $1/2$ accounts 
for the two faces of the disk). The total emitted flux is easily calculated,
however, its spectrum $F_\nu$ is difficult to predict. 
The simplest model assumes that the disk emits as a black body and then one
finds the radial profile of its surface temperature, 
$\Tbb=(F/\sigma)^{1/4}\propto r^{-3/4}S^{1/4}$.
The integrated disk spectrum (multicolor blackbody) has an exponential 
cutoff at $\sim$~keV in X-ray binaries and $\sim 10$~eV in AGN. Such emission
(with the right temperature) is indeed observed in many sources. However,
in addition, one observes a second spectral component that peaks at about 
100~keV.
Example spectra of the famous black hole candidate Cyg~X-1 are shown in
Fig.~3. Most of the time this object spends in the `hard' state when the 
100~keV component is dominant. Sometimes it switches to the `soft' state
where the hard emission is weak and the soft blackbody component is strong.

\begin{figure}[t]
\center{\epsfig{file=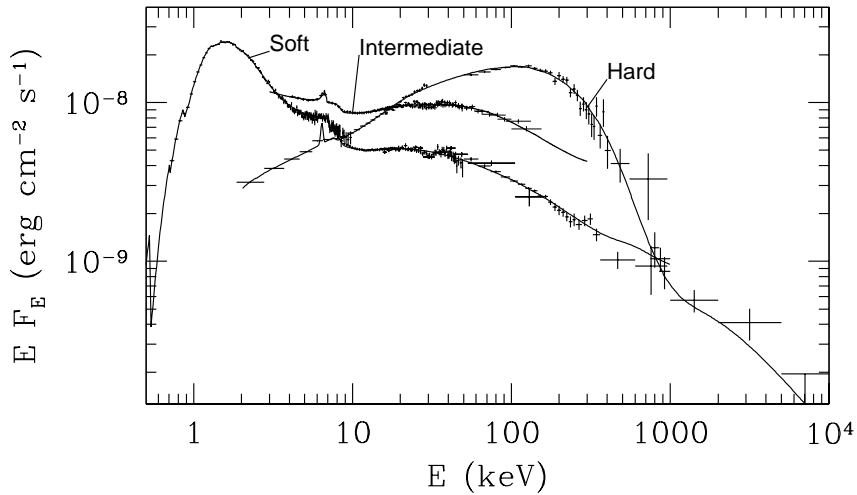,height=6.7cm}}
\caption{
Spectrum of Cyg~X-1 in the hard, soft, and intermediate states.
The hard-state spectrum is obtained in 1991 by CGRO 
(Gierli{\'n}ski {\it et al.}~$^{22}$.
McConnell {\it et al.}~$^{23}$. 
The intermediate-state spectrum was obtained by RXTE on May 23, 1996. 
The soft-state spectrum was obtained by ASCA and RXTE on May 30,
1996 and by OSSE between June 14-25, 1996 (Gierli{\'n}ski {\it et al.}~$^{24}$). 
The solid curves show the best fits to the data by the {\sc EQPAIR} models. 
(From Coppi~$^{24}$.)
\label{fig:states}
}
\end{figure}

Phenomenologically, a two-peak emission spectrum is well explained. 
There must be two gaseous components that are heated in the accretion flow: 
one, dense, emits as a black body (presumably an optically thick accretion 
disk), and the other one is rarefied and hot, with $kT\sim 100$~keV.
A fraction of blackbody photons from the dense component  
get upscattered in the hot plasma, become hard X-rays,
and form the second spectral peak (e.g. Poutanen~\cite{Poutanen}, 
Zdziarski~\cite{Z}).
This Comptonization process is well understood and one can derive the optical
depth $\tT$ (to electron scattering) of the hot plasma from the spectral data. 
Interestingly, one finds $\tT\approx 1$ and $kT\approx 100$~keV in various
sources which include both X-ray binaries and AGN. However the geometry of the
hot+cold gas is still unknown and it is unclear why the accretion flow chooses 
to heat gas with $\tT\approx 1$. Two basic suggestions are: 
(1) buoyant magnetic loops are generated in the turbulent disk 
and form an active corona (Galeev, Rosner \& Vaiana~\cite{Gal}), and 
(2) the cold disk is transformed into a hot two-temperature torus at a radius 
$r_{\rm tr}$; the electron temperature in the torus is $kT_e\approx 100$~keV
(Shapiro, Lightman \& Eardley~\cite{ShapLE}). 
The two modifications of the standard disk look likely
(see Beloborodov~\cite{B} for a review), yet both are strongly dependent on the 
poorly understood viscosity/heating mechanism.

At small accretion rates, the viscous disk may switch to the advective regime 
where the released heat is transported into the hole before it can
be radiated away. The hot advective disks have low luminosities, much below
$\dM c^2$, and can apply to weak X-ray sources (Menou, Blandford, these
proceedings).


\section{Mini-Disk}
As it was discussed in Sect.~2, the estimated angular momentum of the wind-fed 
accretion flows in X-ray binaries is marginally enough to form a disk. 
Is a disk actually formed there? --- Probably yes, because otherwise accretion 
would occur in the quasi-spherical regime with little dissipation, which is hard 
to reconcile with the observed high luminosities. If so, at what radius $r_0$ 
does the disk form and how does it compare with $3r_g$ (the inner edge of the
standard disk)? What is the critical radius $\rcr$ for the {\it viscous} disk 
formation? 

The critical radius turns out quite large, $\rcr\approx 14r_g$, and the accretion
regime with $r_g<r_0<\rcr$ needs to be addressed (Beloborodov \& 
Illarionov~\cite{BI}).
Such a `mini-disk' is very different from the standard viscous disk 
as regards both the dynamics and the emission mechanism.

\begin{figure}[t]
\center{\epsfig{file=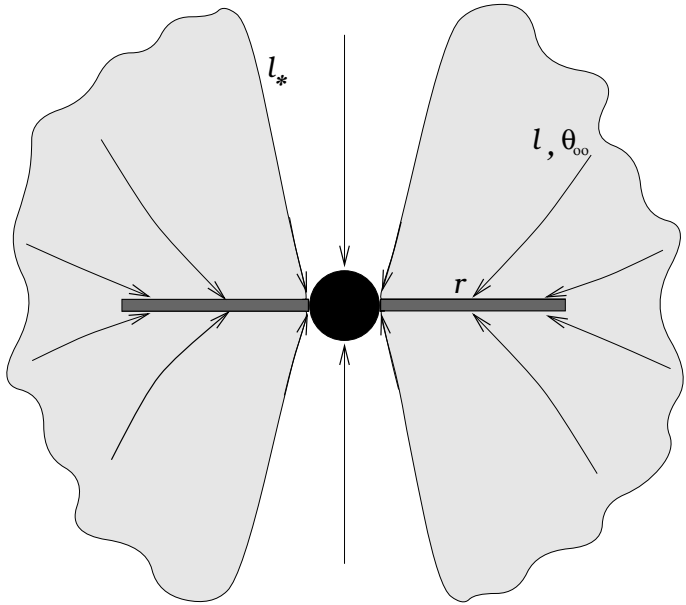,height=8cm}}
\caption{ Schematic picture of the disk formation.
The inflow has angular momentum $l(\thi)$
increasing toward the equatorial plane, $\dd l/\dd\sin\thi>0$.
The collision radius, which is a growing function of $l$, monotonically
increases as $\theta_\infty$ approaches $\pi/2$.
The shadowed parts of the accretion flow collide
outside $r_g$ and form a couple of radiative shocks which sandwich the
caustic. The other part of the flow (at small polar angles where $l<l_*$)
is directly swallowed by the black hole with a low radiative efficiency.
\label{fig:caustic}
}
\end{figure}

\subsection{The Caustic}
Fig.~\ref{fig:caustic}
schematically shows the mini-disk formation in the inner region of a 
quasi-spherical inflow with angular momentum. The inflow is almost radial at 
large radii, it is Compton cooled at $\RC\sim 10^4r_g$ and falls freely toward
the center. Near the black hole, the rotation deflects the infall from the 
radial direction and a disk-like caustic appears in the defocused flow. The flow 
is assumed to be symmetric around the spin axis and also symmetric about the 
equatorial plane. In this plane a ballistic streamline collides with its symmetric 
counterpart. The radius of collision is determined by the angular momentum of
the streamline, $l$,
\be
\label{eq:coll}
   r=\frac{l^2}{GM}-\frac{r_g}{7},
\ee 
where $r_g/7$ is a relativistic correction to the Newtonian formula.
The angular momentum of a streamline depends on its initial polar angle at 
$R\gg r_g$, which we denote $\thi$. A likely distribution $l(\thi)$ has $l=0$ on 
the polar axis ($\thi=0,\pi$) and a maximum $l_0$ at the equatorial 
plane ($\thi=\pi/2$). A simple example is a `solid body' rotation,
$l(\thi)=l_0\sin\thi$. From equation~(\ref{eq:coll}) one can see that streamlines 
with larger $l$ collide at larger $r$.
The polar part of the inflow with $l<l_*=0.75r_gc$ crosses the event
horizon before it can collide, while the equatorial part with $l>l_*$ forms a 
caustic outside $r_g$ (Fig.~\ref{fig:caustic}). The outer edge of the caustic 
is defined by the streamlines with $\thi\rightarrow \pi/2$ and its radius is
$r_0=l_0^2/GM-r_g/7$. The condition for the disk formation outside the 
black hole ($r_0>r_g$) reads $l_0>l_*$.

When the ballistic flow approaches the caustic, it passes through a shock and 
then continues to accrete through the disk. A collisionless shock forms on each 
side of the caustic. The accretion picture shown in Fig.~\ref{fig:caustic} assumes 
that the shock is radiative and pinned to the equatorial plane. This is indeed 
the case if the source luminosity exceeds $10^{-2}\LE$ (Beloborodov \& 
Illarionov~\cite{BI}). The shocked ions pass their energy 
via Coulomb collisions to the radiatively efficient electrons on a timescale
\be
  t_{\rm Coul}=\frac{\sqrt{\pi/2}m_p\theta_e^{3/2}}{m_ec\sT zn\ln\Lambda}
             \approx 17 \frac{T_e^{3/2}}{zn} {\rm s}, 
    \qquad \theta_e\equiv\frac{kT_e}{m_ec^2}.
\ee
Here $n$ is the postshock proton density, $\ln\Lambda\approx 15$ is a Coulomb 
logarithm, and $z>1$ accounts for the possible $e^\pm$ production. The postshock 
electrons have a much smaller temperature than the protons, $T_e\ll T_p$.
They immediately radiate the received heat by Compton scattering, and $T_e$
is controlled by the cooling=heating balance, 
\be
\label{eq:balance}
  \frac{T_e}{\tC}=\frac{T_p}{\tep},
\ee
where $\tC=3m_ec^2/(8\sT F)$ is the Compton cooling time-scale and $F$ is the
local radiation flux. For a radiative shock $F\sim(3/2)nkT_p\vff/\xi$ where 
$\vff=(2GM/r)^{1/2}$ is the free-fall velocity at a radius $r$ and $\xi\geq 4$ 
is the compression at the shock front. Then 
$\tC\approx m_ec^2\xi/(8\sT nkT_p\vff)$ and the energy 
balance~(\ref{eq:balance}) yields
\be
  \frac{kT_e}{m_ec^2}\approx\left(\frac{\ln\Lambda\, m_e}{\sqrt{2\pi}m_p}
                      \frac{\xi c}{4\vff}\right)^{2/5}
          \approx 0.1\left(\frac{\xi c}{4\vff}\right)^{2/5}.
\ee
Density scaled out from this estimate. Given $\xi\approx 4$ and 
$\vff\approx c$ one finds $kT\approx 100$~keV.

The hot postshock layer that radiates most of the proton heat
has optical depth $\tT\approx (\vff/\xi)\tep zn\sT$ and satisfies the relation
$\tT\theta_e\approx 0.1$; one thus gets $\tT\sim 1$ for typical $\theta_e\sim 0.1$. 
In deeper layers, at $\tT>1$, $T_p$ and $T_e$ fall off sharply and the shocked
plasma condenses to a thin dense disk. A similar radiative shock on a surface 
of a neutron star was studied by Zel'dovich \& Shakura~\cite{ZS} and Shapiro \& 
Salpeter~\cite{ShapS}.

The pattern of accretion with shocks pinned to the caustic requires a sufficiently
high accretion rate $\dM$ corresponding to the disk luminosity $L>10^{-2}\LE$.
At smaller $\dM$ the shocked plasma is unable to cool rapidly and it supports 
a quasi-spherical shock front far away from the disk. Such a flow was simulated 
numerically 
by Igumenshchev, Illarionov \& Abramowicz~\cite{IIA}. The simulations show that
the shock tends to shrink toward the disk of radius $l_0^2/GM$ when $\dM$ 
increases.

\subsection{Accretion in the Mini-Disk}
The disk is composed of the shocked matter that entered the disk 
with different horizontal velocities.
A strong turbulence develops under such conditions and efficiently
mixes the disk in the vertical direction.
As a result, the disk shares the horizontal momentum with the feeding infall and
accretes with a vertically-averaged horizontal velocity.
This situation can be described as a `sticky' caustic, 
and it resembles the accretion line in the Bondi-Hoyle-Lyttleton (BHL) problem. 
The infall absorption by the sticky caustic implies energy dissipation, and 
the released heat is radiated away, contributing to the total luminosity.
The process of disk-infall inelastic collision is governed by the corresponding 
law of momentum conservation. The relativistic conservation law 
$\nabla_k T^{k}_{i}=0$ in the Schwarzschild metric yields
\be
\label{eq:radial}
 \frac{\dd u^r}{\dd r}=\frac{\dd\dM}{\dd r}\frac{(\hat{u}^r-u^r)}{\dM(r)}
  -\frac{GM}{r^2u^r}\left(1-\frac{j^2}{\jK^2}\right),
\ee
where $\jK^2(r)=r^2r_gc^2/(2r-3r_g)$ is the squared angular momentum of circular 
Keplerian orbits.
The first term on the right-hand-side is a result of the disk-infall interaction. 
Equation~(\ref{eq:radial}) is similar to the BHL equation~(\ref{eq:bhl}) 
that describes gas motion along the accretion line. The only difference is 
the additional term proportional to $j^2/\jK^2$. It appears because now the
accreting matter rotates and experiences a centrifugal acceleration. 
The repulsive centrifugal force dominates over gravity if $j>\jK$ and it 
can stop accretion. 
Therefore, there exists a critical angular momentum $\jcr$ above which the 
free-fall accretion is stuck and requires viscosity like the standard disk
does. 

\begin{figure}[t]
\center{\epsfig{file=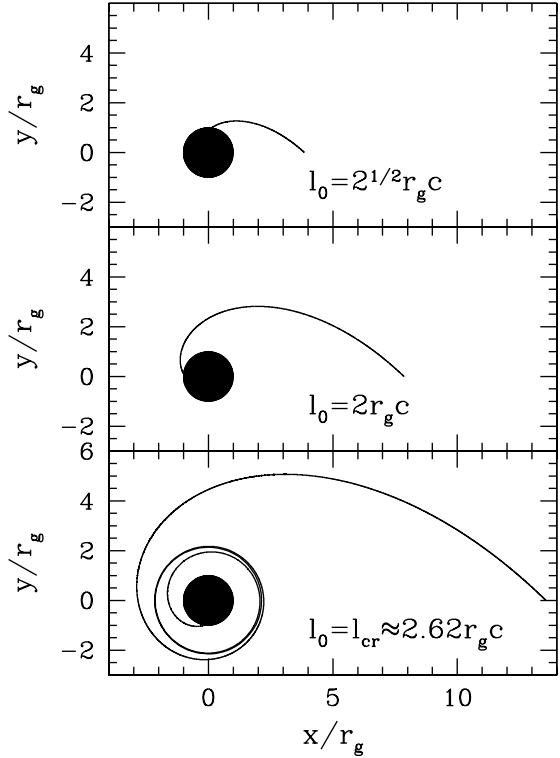,height=10cm}}
\caption{ The trajectory of gas motion in the disk, from the outer edge
into the black hole. In the critical case (bottom panel) the trajectory makes an
infinite number of revolutions at the critical radius $\rcr\approx 2.14r_g$ before
falling into the black hole.
\label{fig:spiral}
}
\end{figure}

The critical angular momentum can be calculated. 
For illustration, consider a flow with $l(\thi)=l_0\sin\thi$; $l_0$ is
the only parameter of the problem. The calculated streamlines of the mini-disk 
are shown in Fig.~\ref{fig:spiral}. A disk with $l_0<\jcr=2.62r_gc$ accretes on 
the free-fall timescale, with negligible horizontal viscous stresses. Its radius 
can be as large as $14 r_g$. A streamline of the disk starts at the outer edge 
$r_0$ with a relatively large angular momentum $j=l_0$ and then, as the streamline 
approaches the black hole, $j$ decreases. This happens because at smaller $r$ the 
disk absorbs the infall with smaller angular momentum;
it allows the disk to overcome the centrifugal barrier 
without horizontal viscous stresses and accrete even with an initially large $l_0$. 
The bottleneck for accretion is at $2r_g$ and here accretion can occur only if 
$j<2r_gc$. One can thus derive a necessary condition for the 
inviscid accretion regime that reads $\bar{l}<2r_gc$ where $\bar{l}$ is the mean
angular momentum of the whole flow.

\begin{figure}[t]
\center{\epsfig{file=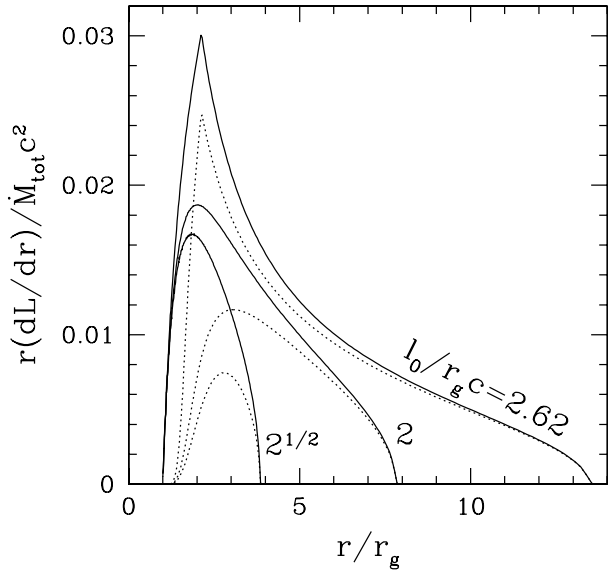,height=8cm}}
\caption{ 
The radial distribution of the disk luminosity: solid curves --
the released luminosity, dotted curves -- the observed luminosity
corrected for the radiation capture by the black hole.
Three models are shown with $l_0=\lcr\approx 2.62 r_gc$,
$l_0=2r_gc$, and $l_0=\sqrt{2}r_gc$.
}
\end{figure}

\subsection{Emission from the Mini-Disk}
The energy dissipated in the inelastic disk-infall interaction is easily 
calculated for a given dynamic solution, like it was done in
the BHL model. The radial distribution of the resulting luminosity
is shown in Fig.~6. It peaks at $2r_g$ where the centrifugal deceleration of the
inflow is strongest (the bottleneck for accretion). 
Here the relative velocity between the disk and the infall
increases and the inelastic collision liberates more energy.
The small size of the main emitting region implies that
the released luminosity is partially captured into the black hole, 
and the corresponding correction is shown in Fig.~6. 

The net radiative efficiency of accretion $\eta=L/{\dM c^2}$ depends on $l_0$: 
it rises from $\eta=0$ at $l_0<l_*$ to a maximum $\eta\approx 0.0372$ at
$l_0=\lcr$. This maximum is comparable to the efficiency of the 
standard disk around a Schwarzschild black hole, $\eta=1-2\sqrt{2}/3\approx 0.0572$.

The energy is released in two steps:
(1) at the shock front where the vertical momentum of the infall is canceled and
(2) inside the disk where the infall inelastically shares its horizontal momentum 
with the disk. The first step heats a postshock layer with $\tT\approx 1$ and 
$kT_e\approx 100$~keV, and the subsequent dissipation in the geometrically thin 
dense disk can emit optically thick cool radiation. The two gaseous components 
thus emit a two-peak radiation spectrum that is typical for black-hole sources.

\subsection{Comparison with Other Accretion Models}
The mini-disk is a sticky caustic in a freely falling flow and in this respect
it resembles the BHL accretion.
In their case, accretion proceeds along a one-dimensional caustic, the accretion
line. The accretion line is fed at each point by matter from
an initially {\it plane-parallel} flow which is axisymmetric and has
{\it zero} net angular momentum. The mini-disk is a two-dimensional
caustic in an asymptotically {\it isotropic} inflow with a {\it non-zero} 
net angular momentum. In contrast to the BHL accretion line,
the radial momentum of matter impinging on the disk is directed inward.
As a result, the mini-disk does not have the degenerate stagnation radius that 
exists in the BHL problem.

For large angular momenta, the disk accretion is stopped by the centrifugal 
barrier and then the mini-disk regime switches to the standard viscous regime. 
The major differences of the mini-disk from its standard counterpart are:
(1) it is inviscid and accretes supersonically,
(2) the energy release in the disk is caused by the infall absorption, and
(3) most of the energy is released inside $3r_g$.


\section{Discussion}
Angular momentum $l$ is a major parameter of an accretion flow.
Two classical modes of accretion --- spherical inflow and viscous disk --- are 
valid in the limits $l\ll r_g c$ and $l\gg r_g c$, respectively. The two limits 
are bridged by an intermediate, but qualitatively different, mini-disk regime.
It takes place for the range of the disk radii $r_g<r_0<14r_g=28GM/c^2$.

Motivation for the theoretical study of the mini-disk comes from the observed 
high-mass X-ray binaries with black holes: the wind-fed accretion flows there 
have $l\sim r_g c$. This fact is interesting: why should wind-fed black holes 
accrete near the threshold for disk formation? One possible explanation is as 
follows. Suppose that the dominant majority of black holes with massive companions 
accrete gas with small $l<l_*$. 
It should be difficult to observe such objects because of their low luminosity.
When $l$ exceeds $l_*$ and the disk forms, the luminosity rises dramatically
and the black hole switches on as an X-ray source. One therefore expects
to see preferentially objects with $l$ above the threshold for disk formation. 
If the $l$-distribution of objects falls steeply toward high $l$, most of the 
bright sources should be near the threshold, i.e. the regime
$\bar{l}\sim r_gc$ can be widespread among observed {\it bright} sources. 

One of the most intriguing properties of accreting black holes is their 
spectral state transitions. An example of such transitions in Cyg~X-1 is 
shown in Fig.~3. The transitions may be caused by modest changes of
the angular momentum of the accretion flow. Such changes should naturally 
occur since the trapped $l$ is very small and sensitive to the fluctuations
of the donor wind. Suppose a black hole accretes in the mini-disk regime and
emits a strong 100~keV emission. If the angular momentum increases
by a factor of 2, the disk has to switch to the standard viscous regime with 
a higher luminosity and softer spectrum. This results in 
a hard-to-soft transition. If the angular momentum 
decreases by a factor of 2, the source luminosity would fall down.
Similar spectral state transitions were recently observed in LMC X-3
(Wilms {\it et al.}~\cite{Wilms}).


\end{document}